\documentstyle[12pt]{article}
\def\np{Nucl. Phys.}
\def\pl{Phys. Lett.}
\def\pr{Phys. Rev.}
\def\prl{Phys. Rev. Lett.}
\def\cmp{Comm. Math. Phys.}

\begin{document}
\baselineskip = 20pt
\input epsf

%
\ifx\epsfbox\UnDeFiNeD\message{(NO epsf.tex, FIGURES WILL BE
IGNORED)}
\def\figin#1{\vskip2in}
\else\message{(FIGURES WILL BE INCLUDED)}\def\figin#1{#1}\fi
\def\ifig#1#2#3{\xdef#1{fig.~\the\figno}
\goodbreak\midinsert\figin{\centerline{#3}}%
\smallskip\centerline{\vbox{\baselineskip12pt
\advance\hsize by -1truein\noindent\footnotefont{\bf
Fig.~\the\figno:} #2}}
\bigskip\endinsert\global\advance\figno by1}

\def\footnotefont{\tenpoint}
\def\figures{\centerline{{\bf Figure
Captions}}\medskip\parindent=40pt%
\def\fig##1##2{\medskip\item{FIG.~##1.  }##2}}
\newwrite\ffile\global\newcount\figno \global\figno=1
\def\fig{fig.~\the\figno\nfig}
\def\nfig#1{\xdef#1{fig.~\the\figno}%
\writedef{#1\leftbracket fig.\noexpand~\the\figno}%
\ifnum\figno=1\immediate\openout\ffile=figs.tmp\fi\chardef\wfile=
\ffile%
\immediate\write\ffile{\noexpand\medskip\noexpand\item{Fig.\
\the\figno. }
\reflabeL{#1\hskip.55in}\pctsign}\global\advance\figno by1\findarg}

\parindent 25pt
\overfullrule=0pt
\tolerance=10000
\def\Re{\rm Re}
\def\Im{\rm Im}
\def\titlestyle#1{\par\begingroup \interlinepenalty=9999
     \fourteenpoint
   \noindent #1\par\endgroup }
\def\tr{{\rm tr}}
\def\Tr{{\rm Tr}}
\def\half{{\textstyle {1 \over 2}}}
\def\calt{{\cal T}}
\def\ie{{\it i.e.}}
\def\np{Nucl. Phys.}
\def\pl{Phys. Lett.}
\def\pr{Phys. Rev.}
\def\prl{Phys. Rev. Lett.}
\def\cmp{Comm. Math. Phys.}
\def\quart{{\textstyle {1 \over 4}}}
\baselineskip=14pt
\pagestyle{empty}
{\hfill CERN-TH/95-221}
\vskip 0.1cm
{\hfill arch-ive/9509171}
\vskip 0.4cm
\centerline{SYMMETRY BREAKING AT ENHANCED SYMMETRY POINTS}
\vskip 1cm
 \centerline{ Michael B.  Green and Michael Gutperle}
\vskip 0.3cm
\centerline{Theory Division, CERN,}
\centerline{CH-1211, Geneva 23, Switzerland}
\centerline{and}
\centerline{DAMTP, Silver Street,}
\centerline{Cambridge CB3 9EW, UK}
\vskip 1.4cm
\centerline{ABSTRACT}
\vskip 0.3cm
The influence of world-sheet boundary condensates on the toroidal
compactification  of bosonic string theories is considered.  At the
special points in the moduli space at which the closed-string theory
possesses an enhanced unbroken
$G\times  G$ symmetry (where $G$ is a semi-simple product of simply
laced groups) a scalar boundary condensate parameterizes the
coset $G\times G/G$.  Fluctuations
around this background define an open-string generalization of the
corresponding chiral nonlinear sigma model.  Tree-level scattering
amplitudes of  on-shell massless states (\lq pions')
reduce to the amplitudes of the principal chiral model  for
the group $G$ in the low energy limit.  Furthermore, the condition
for
the vanishing of the
renormalization group beta function at one loop  results in
the familiar equation  of motion for that model.  The
quantum corrections to the open-string theory generate a mixing of
open and closed strings so that the coset-space pions mix with the
closed-string $G\times G$ gauge fields, resulting in a Higgs-like
breakdown of the symmetry to the diagonal $G$ group.  The case of
non-oriented strings is also discussed.
\vskip 3.3cm
CERN-TH/95-221 \hfil\break\indent
August 1995
\footnotetext
{M.B.Green@damtp.cam.ac.uk
\quad\qquad M.Gutperle@damtp.cam.ac.uk}

\vfill\eject
\pagestyle{plain}
\setcounter{page}{1}

\section{Introduction}

The study of conformal field theory on two-dimensional manifolds with
boundaries has several interesting applications.  The one that has
most relevance to this paper is the description of the perturbation
theory of interacting open and closed bosonic strings, in
which the two-dimensional
manifold describes the world-sheet embedded in $D$-dimensional
space-time.

We will be concerned with the r\^ole of extra massless open-string
states that arise by compactification of open-string theories.  A
generic  toroidal compactification of closed-string theory on a
$d$-dimensional torus possesses $2d$ massless $U(1)$ gauge
potentials, resulting in a  $U(1)^d\times U(1)^d$ gauge symmetry.
These gauge fields are excitations around the condensates of
background metric and antisymmetric tensor fields, $G_{\mu
i}$ and $B_{\mu i}$ (where $i$ labels the compactified directions and
$\mu$ labels the space-time dimensions).
There are also $d^2$ massless scalar closed-string states associated
with the $O(d,d)/(O(d)\times O(d))$ moduli space of toroidal
compactifications. In addition, a theory with open strings has $d$
extra massless scalar states arising from compactification of the
massless open-string vector state --  the excitation of the boundary
condensate of the background open-string vector potential.

The symmetry of the closed-string theory is well-known to be enhanced
on
special sub-manifolds of the background moduli space so it is of
interest to study the spectrum  of the open-string sector  in the
same backgrounds in a theory with world-sheet boundaries. For the
purposes of this paper
we shall consider the special isolated enhanced symmetry points at
which the symmetry is \lq maximal'  which means that it has the form
 $G\times G$, where
$G$ is  a semi-simple product of simply-laced groups ($A,D,E$) of
rank $d$,
although many considerations also generalize to non-maximally
enhanced symmetry points.

The spectrum of the open-string
sector of  a conformal field theory is related to the spectrum of the
closed-string sector by the modular properties of the theory
\cite{cardy}.  Thus,  the partition function for a theory defined on
an annular
world-sheet with boundary conditions  $A$ at $\sigma=0$ and $B$ at
$\sigma=\pi$ can be expressed as the trace over open-string states,
\begin{equation}\label{eq1.2}
Z_{AB}(w)={\rm tr} (w^{H_{AB}})=\sum_iN_{AB}^i\chi_i(w),
\end{equation}
(where $H_{AB}$ is the open-string hamiltonian and
$w=e^{2i\pi\tau'}$, where $2\pi\tau'$ is the imaginary proper  time
around the annulus).  The $\chi_i(w)$ are the conformal
characters of the representation labelled by
$i$ of the Virasoro algebra and $N_{AB}^i$ are integers counting how
often one representation appears in the partition function.
A modular transformation $\tau'\to \tau =  -  1/\tau' $ maps the
annulus to  a
cylinder and the partition function can be expressed as a transition
matrix element between the two boundary end-states states,
\begin{equation}\label{eq1.3}
Z(q)=\langle A\mid  q^{H_{cl}}\mid B\rangle,
\end{equation}
where  $q=e^{i\pi\tau}$  and $H_{cl}$ is the closed-string
hamiltonian.  This demonstrates that the
boundary conditions $A,B,..$ are in one
to one correspondence to the boundary states $\mid A\rangle,\mid
B\rangle,..$.

This relation between the annulus and cylinder will be used  in
section 2 to show that the presence of enhanced affine symmetry at
special points in the  moduli space  of toroidal compactifications of
closed-string theory implies the presence of extra massless
scalar states in the open-string sector.  Vertex operators for the
extra massless open-string and closed-string states are given in
terms of the affine algebra associated with the enhanced symmetry.

Open-string vertex operators arise from fluctuations in background
boundary condensates.  The study of conformal field theories in the
presence of non-trivial boundary condensates (boundary conformal
field theory) has been motivated by several physical applications
including  monopole-catalysed baryon decay
\cite{callan4},\cite{affleck2}, the Kondo effect \cite{affleck1},
tunneling  in quantum wires \cite{affleck3}, dissipative quantum
mechanics \cite{legget},\cite{callan5} and the fractional quantum
Hall effect \cite{fendley}.  In these applications the  boundary
condensate is defined by a marginal boundary term added to the
bulk action,
\begin{equation}\label{pert2}
S=S_0+h_b\int_{\partial\Sigma}d\sigma  \psi_b(\sigma),
\end{equation}
where $\psi_b$ is a boundary operator (and $\sigma$ is the coordinate
tangential to the boundary). If the boundary term is a truly marginal
perturbation to the bulk theory (so that $\psi_b$ has scaling
dimension one) the resulting theory will still be conformally
invariant.  The moduli space of the theory will generally include
parameters that enter into this boundary term.   Several different
cases have
been studied in the literature.  One example
\cite{callan1},\cite{callan1a}, is that
of a  single free compactified boson on the half line with an action,
\begin{equation}\label{eq1.6}
S=\frac{1}{8\pi}\int d\sigma d\tau(\partial_\alpha X)^2-\int d\sigma
\frac{1}{2}(ge^{\frac{i}{\sqrt{2}}X(0,\sigma)}
+\bar{g}e^{\frac{-i}{\sqrt{2}}X(0,\sigma)}),
\end{equation}
where $g$ is a (complex) constant parameter.
Another example is the coupling of a constant antisymmetric
tensor field to the boundary in open string theory \cite{callan2}
\begin{equation}\label{pert3}
S=\frac{1}{8\pi}\int d\sigma d\tau(\partial_\sigma X^\mu )^2+\int
d\sigma
F_{\mu\nu}\partial_\tau X^\mu(0,\sigma) X^\nu(0,\sigma),
\end{equation}
 which is also an exact deformation of the bulk theory.
The case of two free bosons in the presence of both (\ref{eq1.6})
and
(\ref{pert3}) has been discussed in \cite{callan3} and the case of
$n$ free bosons in \cite{yegulalp}. In all cases the exact solution
is given by a modified boundary state $\mid B \rangle_g$   which
generally depends on the parameters  of the perturbation.   However,
at points in the
moduli space of the bulk theory at which there is enhanced symmetry
the boundary conformal field theory may become independent of some
(or all) of these  parameters.  This
was used as important tool  in solving the
boundary scattering problem  in \cite{callan1a} and \cite{pol1} where
the boundary interaction was fermionized and the
theory at the self-dual point was mapped into a theory of free
fermions with twisted boundary
conditions.
In all these cases the condensates were taken to be constant,
i.e.,the parameters are independent of the position on the boundary,
$\sigma$.
[Alternatively, the bulk degrees of freedom can be integrated out and
the boundary
conformal field theory is then expressed as a 1-dimensional
reparameterization invariant field theory \cite{callan6} (an approach
that makes contact with
dissipative quantum mechanics \cite{callan5} and has also been used
to discuss perturbations like (\ref{eq1.6}) and (\ref{pert3})
\cite{freed}),\cite{cff1}.]

The more general situation in which the boundary term is not marginal
is of importance in defining off-shell string field theory in a
background independent manner and is the subject of several
interesting papers
\cite{wittenrooo},\cite{wittenroo},\cite{shatione}.

In section 3 we will argue that, at the enhanced symmetry points of
the toroidally compactified theory  constant scalar boundary
condensates (such as that in (\ref{eq1.6})  and generalizations)
define coset spaces $G\times G/G$ and open-string vertex operators
attached to the boundary describe excitations around these
backgrounds.
The tree-level  S-matrix of the open-string
sector will be described and the nonlinear symmetries associated
with this coset space model  (which is equivalent to the
principal chiral model for the group $G$) will be demonstrated.

The connection between the open-string  tree amplitudes and the
nonlinear sigma model
is made precise in section 4, where the low-energy limit of the
string theory is considered.  Firstly,  the limit of some elementary
S-matrix elements is discussed and shown to correspond to low-order
terms in the expansion of the S-matrix of the sigma model.  The
one-loop beta function for the theory with an arbitrary scalar
boundary condensate is then considered.  The lowest-order condition
for conformal invariance (the vanishing of the one-loop  beta
function) is shown to coincide with the equation of motion for the
sigma model.   Higher-order corrections correspond to higher
derivative nonlinear sigma models (which also couple to the field
strength of
the massless open-string \lq photon').

The interaction between open and closed strings will be considered in
section 5.  This interaction arises naturally from string
perturbation theory and it results in a significant mixing between
the open
and closed sectors.  In the uncompactified theory this leads to a
Higgs-like mechanism in which the antisymmetric tensor potential in
the closed-string sector gains a
mass by absorbing the neutral massless vector potential (the \lq
photon') of the open-string sector.  This was pointed out a long time
ago \cite{cs1} and was investigated in more detail in
\cite{st1},\cite{gw1}.
In the generalization to generic
$d$-dimensional  toroidal compactifications  $d$
linear combinations of the $2d$ $U(1)$ gauge fields gain mass
by absorbing the $d$  extra massless open-string scalar states.  The
generic $U(1)^d\times U(1)^d$ gauge
symmetry is thereby reduced to a diagonal $U(1)^d$.  This is
a stringy generalization of the abelian Higgs effect. We shall
demonstrate that the non-abelian generalization of this
effect at enhanced symmetry points.  The extra  massless
scalar open-string states in the
coset space $G\times G/G$ are eaten by dim($G$) of the generators of
the $G\times G$ enhanced symmetry, giving a non-zero mass to half of
the gauge bosons and leaving the diagonal $G$ symmetry unbroken.
This should be contrasted with the effect in purely closed-string
theories where the enhanced gauge symmetries are broken by marginal
perturbations away from the enhanced symmetry point, in which case a
subset of the gauge particles become massive by mixing with the
massless closed-string moduli fields.

 The generalization to theories with non-orientable world-sheets  is
considered in section 6.  In this case the closed-string sector only
has the diagonal $G$ symmetry to begin with.  At enhanced symmetry
points there are two distinct
definitions of the projection onto the non-orientable open-string
sector.  In one of
these (which is the same projection as the one used at generic points
in moduli space) the massless open-string scalars lie in the coset
the coset $G/ U(1)^d$ (where $d$ is the rank of $G$).  There is no
mixing between the massless states of the open and closed sectors so
the diagonal $G$ symmetry again remains as the unbroken gauge
symmetry.  At the enhanced symmetry points a
generalization of this projection is more natural in the fermionic
formulation of the theory,  which eliminates all
the massless open-string scalar states, again leaving an unbroken
diagonal $G$ gauge symmetry.

\section{Enhanced symmetry points}
A striking feature of toroidal compactifications of
closed-string theory is the occurrence of enhanced gauge
symmetry for special values of the moduli of the target-space torus.
This section will highlight a corresponding enhancement of symmetry
at the same values of the moduli  in the open-string sector of the
theory in which there are both open and closed strings.

\subsection{Review of closed-string sector}
The $U(1)^d\times U(1)^d$  world-sheet affine algebra associated with
generic compactifications of $d$ free bosons  gives rise, in
closed-string theory, to target-space gauge fields which are  the
Kaluza-Klein modes of the antisymmetric tensor and metric fields.
The $d$-dimensional toroidal compactification of the target space on
the
lattice $\Lambda=2\pi Z^d$  is encoded in the background fields $G,B$
which
parameterize the moduli space of toroidal compactifications
$O(d,d;R)/ (O(d,R)\times O(d,R)) $  \cite{narain}.   Each modulus is
a  massless
scalar closed-string state
so the generic number of massless scalars in the closed-string sector
is the dimensionality of the
moduli space, $d^2$. The action containing the constant
background fields is given by,
\begin{equation}\label{res}
S_d=\frac{1}{4\pi\alpha'}\int
d^2z \partial_\alpha X^i\partial_\beta X^j(\eta^{\alpha \beta}G_{ij}+
\epsilon^{\alpha \beta}B_{ij})
\end{equation}
(reparametrization ghosts are incorporated  in the standard manner
and will not be discussed explicitly here).
The term in the action containing the $B$ field is a total
world-sheet derivative
and  shifts the left and right
moving momenta $p,\bar p$ in the compactified dimensions,
\begin{equation}\label{momenta}
p_{i}  =  (\Pi_i+(G-B)_{ij}L^j)\qquad
\bar{p}_{i} = (\Pi_i- (G+B)_{ij}L^j).
\end{equation}
Here $L^i$ is in the lattice $\Lambda$ while $\Pi_i$ denotes the
canonical momentum which takes values on the
dual lattice,  $\Lambda^*=Z^d$ (and  the choice  $\alpha'=1$ has been
made for  convenience). The physical state conditions are,
\begin{equation}\label{physstate}(L_0 -1)|\Phi\rangle= 0 = (\tilde
L_0
-1) |\Phi\rangle,\end{equation}
where the zero Virasoro modes are defined by,
\begin{equation}\label{Lo}
L_0 =\frac{1}{4}k^2+\frac{1}{4}p^2+N, \qquad
\tilde{ L}_0 = \frac{1}{4}k^2+\frac{1}{4}\bar{p}^2+\tilde{ N},
\end{equation}
($k^\mu$ denotes the uncompactified momentum in the $26-d$
dimensions and $N$, $\tilde  N$ are the level numbers in the
left-moving and right-moving sectors).
The closed string hamiltonian $L_0+\tilde{ L}_0$ is given by
inserting (\ref{momenta}) into ({\ref{Lo}),
\begin{equation}\label{rteg}
L_0+\tilde{ L}_0=\frac{1}{2}k^2+N+\tilde{ N}
+\frac{1}{2}\Pi_iG^{ij}\Pi_j+\frac{1}{2}L^i(G-BG^{-1}B)_{ij}L^j
+L^iB_{ij}G^{jk}\Pi_k.
\end{equation}

The ocurrence of extra massless vector states at special values of
the internal momenta signals the enhanced gauge symmetry, which is
associated with an enhanced world-sheet affine algebra.  If one
dimension is compactified to the self-dual radius that is the fixed
point of the T-duality transformation, $R\to 1/R$,  the enhanced
symmetry is $SU(2)\times SU(2)$.
Higher-dimensional  toroidal compactifications are best described in
terms of
constant non-trivial
background fields, $G_{ij}$ and $B_{ij}$, and the maximally enhanced
symmetry
points are generalized fixed points of the $T$-duality group,
$O(d,d;Z)$.  If
$B=0$ the enhanced symmetry   $SU(2)^d\times SU(2)^d$ arises but more
generally semi-simple products of simply-laced
Lie algebras (i.e., of type $A,D,E$) of total rank $d$ can be
obtained.   The maximal  symmetry $G\times G$  (where $G$ is a
simply-laced Lie algebra of rank $d$) is achieved by choosing the
background fields in the following way \cite{tsd}.  Let $C_{ij}$
denote the Cartan matrix of the simply-laced Lie algebra
$G$ and define the background fields as
$G_{ij}=1/2C_{ij}$ and $B_{ij}=G_{ij}$ for $i>j$ , $B_{ij}=-G_{ij}$
for $i<j$ and $B_{ii}=0$.
With these choices  $E \equiv G+B\in SL(d,Z)$. The points in the
compactification moduli space with maximal enhanced gauge symmetry
are
generalized fixed points under the duality transformation $O(d,d;Z)$
generated by a combination of $SL(d,Z)$ conjugation by a matrix $M$
and a shift of $B_{ij}$ by an antisymmetric integer-valued matrix,
$\Theta$.  This transforms $E$ to
$E^{\prime}=M^t(E+\Theta)M$.  For example, in the special case with
$M=E^{-1}$ and $\theta= E^t - E$, duality reduces to $E\to E' =
E^{-1}$.  The extra gauge potentials that arise at points of
enhanced symmetry are labelled by roots in the Lie algebra
lattice of $G$.

The condition that there is an enhancement of the gauge symmetry
(i.e., extra  massless closed-string vector states) requires either
$p=0$, $\bar{p}^2=4$, $N=1$ and $\tilde N=0$  or  $\bar{p}=0$,
$p^2=4$,
$N=0$ and $\tilde N=1$.  In either case it  follows from (\ref{Lo})
and
 (\ref{momenta}) that the condition for enhanced symmetry in the
closed-string sector is
\begin{equation}\label{moreclosed}
L^iG_{ij}L^j=1.
\end{equation}

In addition to the enhancement of the number of massless gauge
particles  there are $({\rm dim}G)^2$ extra massless closed-string
scalar states at a maximally enhanced symmetry point with symmetry
$G\times G$.  As is well understood, if the theory is deformed away
from such an enhanced symmetry point $4d^2 -4d$  of these scalars are
eaten by the extra massless vectors which then become massive and the
rest of the scalars also gain mass apart from the $d^2$ that remain
massless at generic compactifications.  In other words, the
enhanced symmetry is spontaneously broken by deformations away from
the enhanced symmetry point.  The main focus in this paper is on a
mechanism that operates in the presence of boundaries that breaks the
enhanced symmetry without this deformation.

\subsection{Boundary states in the absence of a boundary condensate}
Since the constant $B_{ij}$ term in the action is a total derivative
it gives a boundary term in theories with world-sheet boundaries,
\begin{equation}\label{boundterm}
S_B=\frac{1}{4\pi}\int_{\Sigma}{d^2\sigma
\epsilon^{\alpha \beta }\partial_\alpha X^{i}\partial_\beta
X^{j}B_{ij}} = \frac{1}{4\pi}\oint_{\partial\Sigma} d\sigma
\partial_\sigma
X^{i}B_{ij}
X^{j},
\end{equation}
where $\partial_\sigma $ denotes the tangential derivative on the
boundary.
The term (\ref{boundterm}) is equivalent to coupling a constant
background magnetic field to the boundaries of a neutral open string
(a string with charges of equal magnitude and opposite sign at each
end).
The string coordinate $X^i$ now satisfies the free boundary
conditions,
$(G_{ij}\partial_\tau X^j+B_{ij}\partial_\sigma
X^j)|_{\partial_{\Sigma}}=0$, which guarantee  the conservation of
the world-sheet energy-momentum ($T(z) = \bar T(\bar z)$ when the
world-sheet is the upper-half plane).  If
the world-sheet with a boundary is parameterized as a semi-infinite
cylinder these boundary conditions  are operator conditions on the
boundary state at the end of the cylinder,
i.e.,
\begin{equation}\label{bc}
(G_{ij}\partial_{\tau}X^{i}+ B_{ij}\partial_{\sigma}X^j)\mid
B\rangle=0,
\end{equation}
and the modes of the energy-momentum tensor satisfy
\begin{equation}\label{viramodes}
(L_n-\tilde{ L}_{-n})\mid B\rangle=0,
\end{equation}
where $L_n$ and $\tilde L_n$ are the Virasoro modes.

The boundary state in the cylinder frame that satisfies (\ref{bc})
can be constructed in terms of the modes of the closed-string
coordinates, $\alpha_n^i$ and $\tilde \alpha_n^i$. The state is
given
by \cite{callanx},\cite{pol2},\cite{callan2},
\begin{equation}\label{3}
\mid B\rangle =
C\exp(-\sum_{n>0}\frac{1}{n}\alpha_{-n}^{i}M_{ij}\tilde{
\alpha}_{-n}^{j})
\sum_{L\in Z^d}\mid
p_i ,\bar{p}_i \rangle
\end{equation}
(again ignoring  the terms containing  the reparametrization ghosts)
where,
\begin{equation}\label{pdef}
p_i=(G-B)_{ij}L^j,\qquad \bar{p}_i=-(G+B)_{ij}L^j ,
\end{equation}
and
\begin{equation}\label{mdef}
M^i_{\ j}=[(G+B)(G-B)^{-1}]^i_{\ j}
\end{equation}
is an element of $SO(d,R)$.
The effect of the boundary term is a rotation in the
compactified space of the left-movers with respect to the
right-movers,
\begin{equation}\label{bc5}
(X^i(\sigma)-{M^i}_j\bar{X}^j(-\sigma))\mid B\rangle=0.
\end{equation}
The sum over the
lattice vectors ensures that the total canonical momentum
entering or leaving the boundary vanishes, i.e. $\Pi_i=0$.
The normalization constant $C$  of the boundary state can be
determined by
exploiting the modular properties of the one-loop open-string
partition function as will be seen below.

In the case of non-orientable open-string theories  a general
world-sheet is a Riemann surface with an
arbitrary number of cross-caps in addition to boundaries and handles.
 A cross-cap can be represented by a state, $|C\rangle$, at the end
of a cylinder on which the coordinates satisfy $\left(X(\sigma) -
X(\pi + \sigma)\right)|C\rangle = 0$ and  $\left(\partial_\tau
X(\sigma) +
\partial_\tau X(\pi + \sigma)\right)|C\rangle = 0$.

\subsection{Massless open-string states and modular transformations}
We  shall now turn to
consider the spectrum of open strings. The internal non-abelian
symmetry will come from compactification rather than from Chan--Paton
factors at the string end-points and since  we are dealing with the
bosonic theory the spectrum will inevitably contain a tachyon and a
massless gauge particle -- a \lq photon'. For most of the paper we
shall be considering the orientable open-string theory in which
non-orientable world-sheets (such as M\"obius strips and  Klein
bottles) are eliminated by including a rather trivial $U(1)$
Chan--Paton factor.  Although the photon is present in
this case all states are neutral under its  $U(1)$ and it will
decouple in the low energy limit.   In the absence of any
Chan--Paton factors the theory is non-orientable and the photon
is absent.

At particular points in moduli space, the space of $B$ and $G$,
extra massless open-string scalar states arise
at precisely the same points as the enhanced symmetry points
of the closed-string sector.  This can be seen in a particularly
direct manner by considering modular transformations of the annulus
diagram.  On the one hand this diagram may be viewed as a loop of
open string while it may also be viewed as a cylindrical world-sheet
which describes a closed string propagating between an initial and a
final
boundary state.   As described in the introduction these two
descriptions are related by a modular
transformation that thereby relates the spectrum of the open-string
sector to that of the closed string.

The annulus diagram (the one-loop partition function for an open
string)  with background $B$ and $G$ fields and is given by the trace
over open-string states,
\begin{equation}\label{openpart}
Z(B,G) = \int {dt \over t } Z_{op}(w)=  \int {dt' \over
t' }{\rm tr} (w^{L^{op}_o-1})=  \int{dt' \over t' }\int d^D p
\sum_{k=0}^\infty N_k w^{k + p^2 -1} \end{equation}
where $D=26-d$, $L_0^{open}$ is the open-string hamiltonian,
$t'=-i\tau'$
is the proper (euclidean) time in
the open-string channel and  $w=e^{2i\pi t'}$.  $N_k$ is
the number of open-string states with mass $m^2=k -1$.
This function $Z$ can also be calculated by expressing the partition
function as the matrix element of a closed string propagating between
two boundary states,
\begin{eqnarray}\label{closedpart}
Z (B,G) & =& \int dtZ_{cl}(q) = \int dt\langle B\mid
q^{L_0+\tilde{ L}_0-2}\mid
B\rangle \nonumber\\
&= &  C^2\int dt
q^{-2}\prod_{n=1}^\infty(1-q^{2n})^{-24}\sum_{m\in
Z^d}q^{\frac{1}{2}m^i(G-BG^{-1}B)_{ij}m^j},
\end{eqnarray}
where $t =-i\tau$ is the proper (euclidean) time in the cylinder
channel and
$q=\exp(i\pi t)$.  The equality of the two expressions for $Z$ is
made manifest by the change of variables in (\ref{closedpart}) from
$\tau$ to
$\tau^{\prime}=-1/\tau$ which maps the cylinder into the
annulus.  After this change of variables and a Poisson resummation
of
the sum over the winding modes (\ref{closedpart}) becomes,
\begin{eqnarray}\label{modclosed}
&& Z  (B,G) =\int d\tau' Z_{cl}^{\prime}(w)\nonumber\\
          &&=  C^2\int {d \tau' 2^{d/2}\over (i\tau')^{14 -
\frac{d}{2}}}
{\rm det}(G-BG^{-1}B)^{-\frac{1}{2}}w^{-1}\prod_{n=1}^\infty
(1-w^n)^{-24}\sum_{n\in
Z^d}w^{n^i(G-BG^{-1}B)^{-1}_{ij}n^j}.
\end{eqnarray}

Comparison of (\ref{openpart}) and (\ref{modclosed}) determines the
mass spectrum of the open-string states in terms of
the bulk background fields, $G$ and $B$, of the closed-string sector.
 Explicitly, the spectrum of massless open-string states is given by
the $w^0$ terms in  (\ref{closedpart}).  The expansion  of
$(1-w^n)^{-24}$ leads to the usual states of the massless photon and
the $d$ scalars that come from its compactified components.
Additionally,  the condition that there is an enhancement in the
spectrum of massless scalar open strings is seen from the last factor
to be,
\begin{equation}\label{moreopen}
n_i[(G-BG^{-1}B)^{-1}]^{ij}n_j=1.
\end{equation}
This should be compared with the condition (\ref{moreclosed}) for an
enhancement in the closed-string massless spectrum.   Using
$(G-BG^{-1}B)^{-1}=(E^{-1})^tGE^{-1}$, where $E = G+B$ it is evident
that the solutions of equations (\ref{moreopen}) and
(\ref{moreclosed}) are  proportional,
\begin{equation}\label{c3}
n_i=(G+B)_{ij}L^j.
\end{equation}
This shows that the condition on $B$ and $G$ for an enhancement of
the spectrum of  massless open-string scalar states is precisely the
same as the condition for the enhancement of the spectrum of
closed-string massless vector states.  This argument is valid
whenever the symmetry is enhanced (not just at the self-dual points
of maximal symmetry).

The open-string spectrum can also be determined directly as the
spectrum of the open-string hamiltonian,
\begin{equation}\label{openham}
L_0^{open} = \Pi_i \left[(G- B G^{-1} B)^{-1}\right]^{ij}\Pi_j + k^2
+ N ,
\end{equation}
where the canonical momentum, $\Pi_i \in Z^d$, again demonstrating
the extra open-string states satisfying $L_0^{open}$ when
(\ref{moreopen}) is satisfied.

\subsection{Vertex operators for the massless states}

{\it (a)  The bosonic construction}\hfill\break\noindent
The level-one Kac--Moody algebra can be represented in terms of
bosons
by the Frenkel--Kac--Segal construction  of the group $G_L\times
G_R$. The
left-moving currents are given (in the cylinder channel
with $z=e^{i(\tau+\sigma)}$) by,
\begin{equation}\label{vertkacleft}
H^{i}(z)= \partial X^i,\qquad  E^{\lambda}(z) =
c(\lambda):e^{i\lambda^i X^i(z)}:,
\end{equation}
while the right-moving currents are,
\begin{equation}\label{vertkacright}
 \tilde{ H}^{i}(\bar{z}) =M^i_{\ j}\bar{\partial}X^j, \qquad
 \tilde{ E}^{\lambda}(\bar{z}) =
- \tilde c(\lambda):e^{i\lambda^jM_{ij} \bar{X}^i(\bar{z})}:.
\end{equation}
Here $H^i,\tilde{ H}^i$ denote the currents in the Cartan
subalgebra of $G_L$ and $G_R$ respectively. The $E^\lambda$ are
constructed with winding number states and $\lambda$ are roots of
the
Lie algebra of $G$. The cocycles $c(\lambda)$ are constructed in
terms
of
the zero modes of $X$ and $\bar{X}$.  The root lattice for the
right-moving algebra has been rotated with the orthogonal matrix
$M^i_{\ j}$, which leaves the operator product expansions unaltered.
The currents, $J_L^a(\sigma)$, that satisfy the algebra,
\begin{equation}\label{km}
[J_L^a(\sigma),J_L^b(\sigma^\prime)]=f^{abc}J_L^c(\sigma)
\delta(\sigma-\sigma^\prime)+\delta^{ab}\delta^\prime
(\sigma-\sigma^\prime),
\end{equation}
are linear combinations of the left-moving currents $H,E$, where
$f^{abc}$ are the structure constants for the Lie algebra
$G_L$.
The same construction also applies to the right-moving currents that
are functions of $\bar z$.

In situations in which there are affine symmetries the  left-moving
and right-moving  currents  the boundary conditions on the bosonic
fields translate into conditions on the currents,
$J_L^a(z)-  J_R^a(\bar{z}) =0$,  at $z=\bar{z}$ on the half-plane.
Transforming to the semi-infinite cylinder this translates into the
condition on the boundary state,
\begin{equation}\label{kacmodes}
(J^a_{Ln}+  J^a_{R-n})\mid B\rangle=0
\end{equation}
(where the modes are defined by,
$J^a(\sigma)=\sum_nJ^a_ne^{in\sigma}$).
The  plus sign in this equation arises from the  rotation of the
weight-one currents through $\pi/2$ in transforming from the half
plane to the cylinder frame.

The vertex operators for the massless closed-string vector states are
given by,
\begin{equation}\label{currvert}
V^{a,\mu}_k(z)=   J_R^a(\bar z)\partial X^\mu(z) e^{ik\cdot (X(z)
+ \bar X(\bar z))},  \qquad
\tilde{ V}^{a,\mu}_k(z) =  J_L^a(z)\bar{\partial} \bar X^\mu(\bar z)
e^{ik\cdot (X(z) + \bar X(\bar z))}.
\end{equation}
The vertex operators  for the massless
open-string scalar states are attached to the boundary (at, say,
$\tau=0$).  They can be written in the closed-string  channel as,
\begin{equation}\label{vertopen}
S^a_k(\sigma) = J_L^a(\sigma)e^{ik\cdot X(\sigma)} \equiv  \half
\left(J_L^a(\sigma) - J_R^a(\sigma) \right)e^{ik\cdot X(\sigma)},
\end{equation}
where the second expression follows from the fact that the
open-string vertices are attached to the boundary and using
$\left(J^a_L(\sigma) + J_R^a(\sigma) \right)|B\rangle=0$.
The massless vector vertex is given by,
\begin{equation}\label{vecvert}
V_k^\mu (\sigma) = \partial_\sigma X^\mu (\sigma) e^{ik\cdot
X(\sigma)}
\end{equation}
(where $ \sigma $ is again the parameter tangential to the boundary).

\vskip 0.3cm\noindent{\it (b) The fermionic
construction}\hfill\break\noindent
Level-one affine algebras also have well-known
representations in terms of free fermions \cite{gross}. For
simplicity
we will here consider the special case of $SO(2d)\times SO(2d)$ which
is implemented by $2d$ left-moving and $2d$ right-moving Majorana
fermions, $\psi_L^i$ and $ \psi_R^i$ ($i=1,\cdots, 2d$).  These
fermions are described by the bulk world-sheet action, $\int
d^2z (\psi_L^i\bar{\partial}\psi_L^i+\psi_R ^i
\partial\psi_R ^i) $ and the currents satisfying (\ref{km}) can be
written as
bilinears in real left-moving and right-moving fermions,
$\psi_L^i,\psi_R ^i\;\{i=1,..,2d\}$,
\begin{equation}\label{c1}
J_L^a(z)=\psi_L^i(z)T^a_{ij}\psi_L^j(z), \qquad
J_R^a(z)=\psi_R ^i(z)T^a_{ij}\psi_R ^j(z),
\end{equation}
where  $T^a_{ij}$ are the antisymmetric generators of $SO(2d)$.
Quantizing in the cylinder channel the massless states of
the adjoint representation lie  in the sector with antiperiodic
boundary conditions on both fermions (the NS/NS sector).
The closed-string boundary state in this sector  is given by,
\begin{equation}\label{bst}
\mid B\rangle=P_{GSO} \exp \left(i\sum_{n\in
Z+1/2}\psi^i_{L-n}\psi^i_{R-n} \right)\mid B\rangle_0
\end{equation}
(where $|B\rangle_0$ is the boundary state involving the space-time
bosons). The $GSO$ projector is given by $(1 + (-1)^F)(1+(-1)^{\tilde
F})/4$, where $F$ and $\tilde F$ are the fermion number operators in
the left-moving and right-moving sectors.  This projection eliminates
states with odd numbers of $\psi$ or $\tilde \psi$ excitations,
resulting in a theory that is equivalent to that given by the bosonic
description.
The state (\ref{bst})  implements the boundary condition,
\begin{equation}\label{aswej}
(\psi^i_{Ln} \pm i\psi^i_{R-n})\mid B\rangle = 0,
\end{equation}
which is the boundary condition on the closed-string states that
corresponds to the condition at an end-point of an open string,
$\psi_L^i = \pm \psi_R^i$.  The
relative factor of $i$ appears in (\ref{bst}) from the  conformal
transformation that rotates the coordinate frame through $\pi/2$ in
transforming conformal spin-1/2 fields from the open-string to the
closed-string frame.
The boundary condition on the currents,
(\ref{kacmodes}) also follows from this definition of the boundary
state.

The different fermionic spin structures give rise  to different
$SO(2d)$ conjugacy classes
of $SO(2d)$.  The  complete set of states includes the
$GSO$-projected $R/R$ and $N/R$ sectors which lead to spinor and
bi-spinor representations (which do not have massless excitations).

The vertex operators of (\ref{currvert})-(\ref{vecvert}) have obvious
descriptions in terms of the fermion representation of the currents.

\section{Boundary condensates}
\subsection{Constant condensates}
The occurrence of extra massless open-string states at the points in
moduli space at which the gauge symmetry of the closed-string sector
is enhanced indicates the presence of
new marginal boundary operators.  Such operators
come from   states located on
the boundary associated with terms in the action of the
form (in the bosonic formalism),
\begin{equation}\label{cdfgw}
S_B=\int d\sigma(\sum_\lambda g_\lambda e^{i\lambda_i X^i}+\sum_i
g_i\partial_\sigma X^i)\mid_{\tau=0}
\end{equation}
(where $g_\lambda$, $g_i$ are constants), which gives rise to the
enhanced symmetry when $\lambda_i$ takes
special values.  In \cite{yegulalp} it was shown these values arise
when $\hat{\lambda}_i$ are roots of the Lie algebra $G$, where
$\hat{\lambda}_i=(\delta^i_j+M^i_{\ j})\lambda^j$ (and $M^i_{\ j}$
was defined in (\ref{mdef}))  This leads to a
modified  boundary state in the cylinder channel,
$\mid B\rangle_g=\exp(S_B)\mid B\rangle$  as can be seen by turning
the $X^i$ modes into left-movers using the boundary
condition (\ref{bc5}).  The effect
of the boundary state is then simply a rotation with respect to the
left-moving zero modes of the currents,
\begin{equation}\label{modboun}
\mid B\rangle_g=\exp(g_{\hat{\lambda}}
E^{\hat{\lambda}}_0+g_jH^j_0)\mid B\rangle.
\end{equation}
In terms of the currents $J_L^a$ this condition can be written as
\begin{equation}\label{bc6}
\mid B\rangle_g=\exp \sum_a g_a J^a_{L0}  \mid B\rangle.
\end{equation}
It follows that a non-zero condensate changes the boundary
condition
(\ref{kacmodes}) into,
\begin{equation}\label{bc8}
(e^{ g_a J^a_{L0}}J^a_{Ln} e^{- g_aJ^a_{L0}}+ J^a_{R-n})\mid
B\rangle_g =0.
\end{equation}
In the fermionic formulation of the current algebra the
action of boundary condensate (\ref{bc6}) is described by a
change in the boundary conditions on the fermions,
\begin{equation}\label{bc7}
\left((e^{- g_aT^a})_{ij}\psi^j_{Ln}+i\psi^i_{R-n}\right)\mid
B\rangle_g.
\end{equation}
These boundary conditions also follow directly from the fermionic
version of the boundary action on the half plane, $z=\tau+i\sigma$
($\sigma\ge 0$),
\begin{equation}\label{mjar}
S_B=  \int d\sigma
\sum_a\frac{g_a}{2}(\psi_L^iT^a_{ij}\psi_L^j+\psi_R ^i
T^a_{ij}\psi_R ^j),
\end{equation}
as the following argument shows.   Together with the bulk action
this may be written in terms of  left-moving fermions only on the
whole complex
plane by making the standard  identification,
$\psi_R(z)=\psi_L (\bar{z})$.
The boundary is thereby replaced by an interaction located at
$\sigma=0$.
The equation of motion for the left-moving $\psi_L^i$ is then given
by,
\begin{equation}\label{klsw}
(\partial_\tau-\partial_\sigma)\psi_L^i+g_aT^a_{ij}
\psi_L^j\delta(\sigma)=0.
\end{equation}
The delta function may be regularized by replacing it with
$1/2a[\Theta(\sigma-a)\Theta(a-\sigma)]$, which is nonzero
in the small interval $\sigma\in [-a,+a]$ with $a\to 0$ so that for
$\sigma<a$ and $\sigma>a$, $\psi_L$ satisfies the
free equation of motion. In the interval $\sigma\in [-a,+a]$ the
perturbation is just a constant matrix.   Fourier  transforming
$\psi_L(\tau,\sigma)$,
\begin{equation}\label{gwesr}
\psi_L^i(\sigma,\tau)=\int d\nu\psi^i_{L\nu}(\sigma)e^{-i\nu\tau},
\end{equation}
leads to a differential equation,
\begin{equation}\label{sdwI}
\frac{d}{d\sigma}\psi^i_{L\nu}(\sigma)=(-i\nu\delta_{ij}
+\frac{g_b}{2a}T^b_{ij})\psi^j_{L\nu}(\sigma),
\end{equation}
in the interval $\sigma\in [-a,+a]$.  Integrating this between
$\sigma=-a$ and $\sigma=a$ gives
\begin{equation}\label{csui}
\psi_L^i(a)=\exp(i\nu a  + gT)_{ij}\psi_L^j(-a) .
\end{equation}
In the limit $a\to 0$,
\begin{equation}\label{bounferm}
\psi_L^i(0^+,\tau) = (e^{g_aT^a})_{ij}\psi_L^j(0^-,\tau) =
(e^{g_aT^a})_{ij}\psi_R^j(0^+,\tau),
\end{equation}
where  $\sigma=0^\pm$ indicates that the  line $\sigma=0$ is
approached from positive and negative values, respectively.   This
leads to the  relation between left-moving and right-moving fermions
in (\ref{bc7}) (the factor of $i$ was explained earlier) so that the
boundary
interaction can be expressed as  a change of the fermionic boundary
condition.

\subsection{Coset space interpretation of boundary condensate}
 As can be seen
from  (\ref{kacmodes}) or (\ref{bc8})   (for the cases of zero and
nonzero
boundary condensates respectively) only a particular linear
combination of the world-sheet currents is conserved by the boundary.
  The combinations of  the boundary values of the left-moving
and right-moving world-sheet currents ,
\begin{equation}\label{dwera}
 K^a(\sigma)=\half(J^a_L(\sigma)-J^a_R(\sigma)),
\end{equation}
\begin{equation}\label{serw}
J^a(\sigma)= \half(J^a_L(\sigma)+J^a_R(\sigma)),
\end{equation}
satisfy the  commutation relations,
\begin{eqnarray}\label{km1}
[J^a(\sigma),J^b(\sigma^\prime)]
&=& f^{abc}J^c(\sigma)
\delta(\sigma-\sigma^\prime)+\delta^{ab}\delta^\prime
(\sigma-\sigma^\prime), \nonumber \\ \
[J^a(\sigma),K^b(\sigma^\prime)]
&=& f^{abc}K^c(\sigma)
\delta(\sigma-\sigma^\prime), \nonumber\\ \
[K^a(\sigma),K^b(\sigma^\prime)]
&=& f^{abc}J^c(\sigma)
\delta(\sigma-\sigma^\prime)+\delta^{ab}\delta^\prime
(\sigma-\sigma^\prime).
\end{eqnarray}

 It follows from (\ref{kacmodes})  that $J$
vanishes on the boundary state while $K$ generates an infinitesimal
shift of the boundary condensate.   Therefore these transformations
define  a symmetric space,  where $J$ parameterizes the unbroken
group ${\cal H}$ while  $K$ parameterizes the  coset
${\cal G}/{\cal H}$.  In the case of interest here ${\cal
G}=G_L\times G_R$ and ${\cal H}=G_{diagonal}$.    The  manifold of
possible boundary states (parameterized by
the couplings $g_a$)  define equivalent vacua of the theory,
transformed into each other by $K$.
The massless scalar open-string states are fluctuations around a
given
vacuum and are the Goldstone bosons of  the spontaneously broken
symmetry.
We will see later that the presence of a boundary causes a breaking
of the local $G\times G$
symmetry of the closed-string sector of the theory.

\subsection{The S-matrix of the Goldstone bosons}
The Goldstone bosons describing the fluctuations around a given
vacuum state  (a state with a given constant condensate) are the
massless scalar  states of the open string sector described by  the
scalar vertex operators, $S^a_k$, (\ref{vertopen}).  Open-string tree
amplitudes are obtained by functional integration on a world-sheet
that has the topology of a disk, while higher-order diagrams have
extra boundaries as well as handles.
The on-shell amplitude for $n$ massless scalar
open-string states with momenta $k_i$ satisfying $k_i^2=0$ is given
by,
\begin{equation}\label{lofh}
A_n(\Phi_1(k_1),\cdots,\Phi_n(k_n))=\Phi_1^{a_1}\cdots
\Phi_n^{a_n}\langle a_1, k_1\mid  S^{a_2}_{k_2}\Delta\cdots \Delta
S^{a_{n-1}}_{k_{n-1}}\mid a_n,k_n\rangle+\ {\rm perms}.
\end{equation}
where $\Delta$ denotes the open string propagator $(L_0-1)^{-1}$
and $\Phi_r^{a_r}$ are the momentum-space wave functions of the
massless
scalar states.  Each vertex operator, $S_k^a$, is proportional to the
current ${\rm J}^a $ at one endpoint or the other
of the open string $(\sigma=0,\pi)$.   The vertex operator in
(\ref{vertopen}) was
expressed in the cylinder channel while those in (\ref{lofh}) are in
the open-string frame, which leads to a sign change in the
combination of currents in the second expression in (\ref{vertopen}),
which  becomes ${\rm J}^a (\sigma) = \half( J_L^a(\sigma) +
J_R^a(\sigma))$.   There are no Chan--Paton
factors so the expression includes a sum
over all non-cyclic permutations of the external states -- this sums
over all orderings of the vertex operators with equal weight.

Now consider the  four-point function  in more detail.  It is given
by
\begin{equation}\label{foureval}
A_4(s,t,u) =\sum_{perms}\int_0^1 dx\langle 0|S^{a_1}_{k_1}
(0)S^{a_2}_{k_2}
(1)S^{a_3}_{k_3} (x)S^{a_4}_{k_4} (\infty)|0\rangle  =
\sum_{perms}\int_0^1 dx\
{\cal I}(x)\ {\cal J}(x),
\end{equation}
where a sum over inequivalent permutations of the order of the
external particles is indicated by $\sum_{perms}$ and
\begin{equation}\label{expmat}
{\cal I}\  \equiv\ \langle 0|e^{ik_1\cdot X(\infty)}e^{ik_2\cdot
X(1)}e^{ik_3\cdot X(x)}e^{ik_4\cdot X (0)}|0\rangle \ = \
(1-x)^{-t/2}x^{-s/2}
\end{equation}
(the Mandelstam invariants are defined by $s=-(k_1+k_2)^2$,
$t=-(k_2+k_3)^2$ and $u=-(k_1+k_3)^2$) and
\begin{equation}\label{currmat}
{\cal J}\ \equiv \ \langle 0|{\rm J}^a(\infty) {\rm J}^b(1) {\rm
J}^c(x)
{\rm J}^d (0)|0\rangle .
\end{equation}
The evaluation of ${\cal I}$ in (\ref{expmat}) involves standard
algebra.  The current algebra part of the matrix element,
${\cal J}$, can be evaluated using,
\begin{equation}\label{cr}
[{\rm J}_n^a,{\rm J}_m^b]=f^{ab}_c{\rm
J}_{n+m}^c+\frac{k}{2}n\delta_{n+m}\delta^{ab}
\end{equation}
(where the modes of the currents are defined by ${\rm
J}^a(x)=\sum_n{\rm J}_n^ax^{-n-1}$).
 The current matrix
element is given by,
\begin{eqnarray}\label{curragain}
{\cal J}(x) &=&
\sum_{n,m}x^{-m-1}\langle 0|{\rm J}_{1}^a{\rm J}^b_n{\rm J}^c_m{\rm
J}^d_{-1}|0\rangle \nonumber\\
 & =&\frac{k}{2}\{f^{\ ab}_ef^{ecd}x^{-1}(1-x)^{-1}+f^{ac}_ef^{edb}
(1-x)^{-1}\} \nonumber\\ &&
+\frac{k^2}{4}\{\delta^{ab}\delta^{cd}x^{-2}
+\delta^{ac}\delta^{bd}+\delta^{ad}\delta^{bc}(1-x)^{-2}\}
\end{eqnarray}
(the vacuum $|0\rangle $ is annihilated by ${\rm J}^a_n$ for
$n\geq 0$ and the
expectation value of a single current vanishes).  From hereon the
level $k$ is to be set equal to 1.
The relations,
\begin{equation}\label{grouptheory}
\delta^{ab}={\rm tr}(T^aT^b),
\qquad f^{\ ab}_ef^{ecd}={\rm tr}([T^a,T^b][T^c,T^d]),
\end{equation}
will be used to to write the expressions in terms of traces of
products of two and
four generators of the Lie algebra, $G$.

The full amplitude is given by combining (\ref{expmat}) and
(\ref{curragain}) and summing over permutations
with cyclically inequivalent orderings  of the  vertex
operators.  The $x$ integrals are simple
combinations of Euler beta functions and the  result may be expressed
as the sum,
\begin{equation}\label{finalfour}
A_4(s,t,u) =  A^{(1)} + A^{(2)} + {\rm perms},
\end{equation}
where,
\begin{equation}\label{resfourone}
A^{(1)}= {\pi^2 \over 2} {\rm tr} (T^{a_1}T^{a_2}T^{a_3}T^{a_4}) {  u
(\sin  \pi {s\over 2}
+   \sin \pi {t\over 2} +  \sin \pi{u\over 2} )\over  \Gamma({s\over
2}+1) \Gamma({t\over 2}+1)  \Gamma({u\over 2}+1)\sin(\pi {s\over 2})
\sin(\pi {t\over 2})\sin(\pi {u\over 2})}
\end{equation}
and
\begin{equation}\label{resfourtwo}
A^{(2)} = {\rm tr}(T^{a_1}T^{a_2}){\rm
tr}(T^{a_3}T^{a_4})\frac{\Gamma(-{t\over
2}+1)\Gamma(-{u\over 2}+1)}
{4\Gamma({s\over 2}+2)\sin{\pi {s\over 2}}}(\sin{\pi {s\over
2}}+\sin{\pi {t\over 2}}+\sin{\pi{u\over 2}}).
\end{equation}
A sum over cyclic permutations of the external states is indicated.
The presence of the tachyon ground state is seen from the ocurrence
of the poles in $A^{(2)}$ at $s,t,u=-2$.  There are no massless
intermediate poles
in the amplitude, reflecting the fact that there is no trilinear
coupling between the massless scalar states and the photon
decouples.

Another way of calculating the amplitudes is by transforming  to
the cylinder frame and writing the open-string vertex operators  in
terms
of the closed-string variables as in (\ref{vertopen}), $S^a_k(\sigma)
= K^a(\sigma) e^{ik\cdot X(\sigma)}$.  The amplitudes may be packaged
into the
generating functional,
\begin{equation}\label{gencyl}
Z[\Phi]=\langle 0\mid \exp(\int d\sigma \int d^dk \Phi^a(k)
K^a(\sigma)e^{ik\cdot X})\mid B\rangle.
\end{equation}
This gives the same expression for the amplitudes as (\ref{lofh})  as
can be seen by expanding the exponential in (\ref{gencyl}) and using
$K^a(\sigma)|B\rangle \equiv J_L^a(\sigma)|B\rangle$   together with
the
current
algebra of $J_L(\sigma)$.  Once all the $K^a$'s have been replaced by
$J_L^a$'s the boundary state can be replaced by the vacuum state (the
bosonic coordinates work in the standard manner).
More generally, the closed-string bra state in (\ref{gencyl}) may be
replaced by an arbitrary physical closed-string state, $\langle P|$,
to describe the coupling between closed and open strings.

The S-matrix elements for on-shell momenta $k_1,\cdots,k_n$
($k_i^2=0$) are
given by
\begin{equation}\label{ftry}
A_n=\frac{\delta^n Z[\Phi]}{\delta
\Phi^{a_1}(k_1)\cdots\delta\Phi^{a_n}(k_n)},
\end{equation}
giving,
\begin{equation}\label{gth}
A_n=\Phi_1^{a_1}\cdots\Phi^{a_n}_n\langle 0\mid \int d\sigma_1
K^{a_1}(\sigma_1)e^{ik_1\cdot X(\sigma_1)}\cdots \int d\sigma_n
K^{a_n}(\sigma_n)e^{ik_n\cdot X(\sigma_n)}\mid B\rangle.
\end{equation}
This expression can be written with a particular ordering of the
operators along the axis of the cylinder even though the open-string
vertices are all attached to the boundary.    In other words, the
$\sigma$ contours have been displaced by infinitesimal amounts along
the cylinder axis to avoid the collision of vertex operators.   A
precise connection with the previous calculation in the open-string
frame involves  averaging over the ordering of these operators along
the axis of the cylinder.    This is equivalent to the principal part
 prescription implicitly used in evaluating   (\ref{foureval}).
Actually, the
result does not generally depend on this ordering  since the
difference between two orderings is a contact term, in which two
vertex operators are evaluated at coincident points, that gives
vanishing contribution for suitably defined generic external momenta.
  This method of displacing the integration contours infinitesimally
away from the boundary is useful not only  because it facilitates the
operator algebra but it also
acts as an ultraviolet regulator.

\subsection{$G\times G$ symmetry of the $S$-matrix}
We shall now investigate the global symmetries of the S-matrix
elements using the formalism based on the cylinder
channel operators in which the vertices are proportional to  $K^a$.
The first symmetry is a linear symmetry under $G$ transformations
generated by $J_0^a$.  This follows from the identity,
\begin{equation}\label{derw}
A_n = \Phi_1^{a_1}\cdots\Phi^{a_n}_n\langle 0\mid \int
d\sigma_1 K^{a_1}(\sigma_1)e^{ik_1\cdot X(\sigma_1)}\cdots \int
d\sigma_n
K^{a_n}(\sigma_n)e^{ik_n\cdot X(\sigma_n)}(1+\epsilon^a J^a_0)\mid
B\rangle,
\end{equation}
which is a simple consequence of the fact that  $J_0|B\rangle =0$.
Commuting the factor $J_0^a$ to the left until it annihilates
the closed-string ground state and making use of  (\ref{km1}) gives
an
$n$-particle amplitude with vertex operators transformed so that they
have the form (to lowest order in $\epsilon^a$),
\begin{equation}\label{dkoi}
\int d\sigma_i ( \Phi^a + f^{abc}\epsilon^b\Phi^c)
K^{a}(\sigma_i)e^{ik\cdot X(\sigma_i)}  .
\end{equation}
This  means that the n-point function is invariant under
transformations of the wave functions,
\begin{equation}\label{jtrans}
\Phi^a \to \Phi^{a\prime}=\Phi^a+f^{abc}\epsilon^b\Phi^c,
\end{equation}
which shows that $J$
acts linearly on the wave functions and is a rotation in the
unbroken
group ${\cal H} = G$.

The $S$-matrix is also invariant under  an infinitesimal modification
of the boundary state,
$\mid
B\rangle_\eta   =(1+\eta^a K^a_0)\mid B\rangle$, which corresponds
to an insertion of a zero momentum open string scalar state with wave
function $\eta^a$.
The change in the $n$-particle amplitude is given by,
\begin{equation}\label{dre}
\delta A_n = \Phi_1^{a_1}\cdots\Phi^{a_n}_n\langle 0\mid
\int d\sigma_1 K^{a_1}(\sigma_1)e^{ik_1\cdot X(\sigma_1)}\cdots \int
d\sigma_n K^{a_n}(\sigma_n)e^{ik_n\cdot X(\sigma_n)} \eta^a K^a_0
\mid
B\rangle.
\end{equation}
Commuting $K_0$ to the left until it annihilates the closed
string vacuum (which is a singlet under $J_{L0}$ and $J_{R0}$
separately)
and using the
commutation relations for $K_0$ that follow from (\ref{km1}) leaves
terms
containing the product of vertices with one vertex replaced by,
\begin{equation}\label{dloi}
\int d\sigma_i f^{abc}\eta^b \Phi^c
J^{a}(\sigma_i)e^{ik\cdot X(\sigma_i)}.
\end{equation}
Each of these factors can then be commuted to the right until the
current density $J^a(\sigma)$ annihilates the boundary state.  This
leaves
terms containing commutators of the form,
\begin{equation}\label{residcurr}
 \left[\int  d\sigma_i f^{abc}\eta^b \Phi^c J^{a}(\sigma_i),\int
d\sigma_j \Phi^d K^{d}(\sigma_j) \right]
=  \int d\sigma_i\int d\sigma_j \eta^b \Phi^c\Phi^d
f^{abc} f^{ade}K^e(\sigma_j)\delta(\sigma_i - \sigma_j)
\end{equation}
(the Schwinger term vanishes because of the antisymmetry of the
structure
constants $f^{abc}$). The contribution from the right-hand side
vanishes for
amplitudes in which the external states have generic momenta.  This
follows by the usual reasoning that the   $e^{ik\cdot X}$ terms in
${\cal I}$ give rise to  factors of
$|e^{i\sigma_i}-e^{i\sigma_j}|^{k_ik_j}$ which multiplies the $\delta
(\sigma_1-\sigma_2)$ from (\ref{residcurr}), thereby vanishing (for
suitably defined $k_i\cdot k_j$).  This means that all the
commutators of  the form (\ref{residcurr})
vanish by analytic continuation to physical momenta.

Thus, the scattering  amplitude with  one zero-momentum Goldstone
boson vanishes, which demonstrates the shift symmetry,
\begin{equation}\label{nonlinsymm}
\Phi^a(x) \to \Phi^{a \prime} (x) = \Phi^a(x) + \eta^a,
\end{equation}
(where $\eta^a$ is a constant) characteristic of the
nonlinearly realized ${\cal G}/{\cal H}$ symmetry of coset-space
models.

Although (\ref{nonlinsymm}) is the symmetry of the S matrix for
generic external momenta, the set of nonlinear field transformations
seen in the
sigma model lagrangian field theory have their counterparts in the
S-matrix in processes with non-generic
external momenta  such that $k_i \cdot k_j=0$.  Here extra
contributions arise from the possibility of  on-shell intermediate
states with $(k_i+k_j)^2=0$.  In the presence of such momentum
configurations the $\delta(\sigma_i-\sigma_j)$ factor in
(\ref{residcurr}) gives non-vanishing contact terms.  In this case
the insertion of a zero-momentum boundary scalar state in an
$n$-particle amplitude is related to an $n-1$-particle
amplitude with a nonlinear transformation on a wave-function,
$\delta\Phi^e=\eta^b\Phi^c\Phi^df^{abc}f^{ade}$.  A more general
discussion of the nonlinear symmetries of the model will be given at
the end of the next section.

\subsection{Amplitudes with massless photons}

The massless vector open-string state does not arise as an
intermediate state in the four-particle amplitude since all states
are neutral under its (Chan-Paton) $U(1)$ charge.
Actually, all open-string amplitudes with a single external photon
vanish so that none of the massless scalar tree diagrams have
intermediate photon poles.  However, amplitudes with two or more
external photons are nonvanishing. For example, the amplitude with
two
photons with polarizations $\xi^{\mu_r}_r(k_r)$ and two massless
scalar states is given by,
\begin{equation}\label{photonscalar}
A_{\gamma\gamma}(s,t,u) =\int \prod_{r=1}^4 dx_r
\mu(x)\xi_{1\mu}(k_1)\xi_{2\nu}(k_2)  \Phi^a_3 \Phi^b_4   \langle
0\mid  V_{k_1}^\mu(x_1)  V_{k_2}^\nu(x_2)  S^a_{k_3}(x_3)
S^b_{k_4}(x_4) \mid 0\rangle,
\end{equation}
where the vertex operator  for a transverse photon  was defined in
(\ref{vecvert}) and  $\mu$ denotes the  Moebius-invariant measure
that can be used to fix three of the $x_r$.  The amplitude can be
written as
\begin{equation}\label{photon2}
A_{\gamma\gamma}(s,t,u)=\int
\prod_{i=1}^{4}dx_i\mu(x)\frac{\xi_1\cdot
\xi_2}{(x_1-x_2)^2}\frac{\Phi_3^a\Phi_4^a}{(x_3-x_4)^2}
\prod_{i<j}(x_i-x_j)^{k_ik_j}+\cdots.
\end{equation}
Only the $\xi_1 \cdot \xi_2$ term is displayed and the  $+\cdots$
refers to other contractions, $\xi_1\cdot k_3\ \xi_2\cdot k_4$ etc.
(which  are determined by  gauge invariance). Evaluating the integral
and summing over all orderings on the disk gives,
\begin{equation}\label{photon3}
A_{\gamma\gamma}=\xi_1\cdot \xi_2
\Phi_3^a\Phi_4^a\frac{\Gamma(-u/2+1)\Gamma(-t/2+1)}{\Gamma(s/2+2)\sin
(\pi s/2)}\left\{\sin (\pi s/2)+\sin (\pi t/2)+\sin (\pi u/2)\right\}
+\cdots.
\end{equation}

\section{Low energy limit and principal chiral models}
Since the tree-level $S$ matrix for the massless scalars has
the symmetries associated with coset space nonlinear sigma models
its low-energy limit should simply be such a model.
In order to make this explicit we shall now consider the low energy
properties of the theory with boundary condensates.  This is a
description in terms of the
massless degrees of freedom at energies such that all the massive
string states are effectively integrated out.

Recall that the  coset model with ${\cal G}=G\times G$ and ${\cal H}
= G$
is equivalent to the principal chiral sigma model in which the scalar
fields parameterize the group manifold of $G$.  The Goldstone boson
(\lq pion') fields that arise in principal chiral
models    parameterize the group manifold $G$, so that a group
element is given by
$g(X)=\exp(\Phi^a(X)T^a)$. The action of the principal chiral model,
\begin{equation}\label{chiralact}
\int d^D x \half {\rm tr}(g^{-1}\partial_\mu g g^{-1}\partial^\mu g),
\end{equation}
has global invariance under left and right multiplication with
group elements $\rho_L$ and $\rho_R$.

 We will first indicate  how the low-energy effective action can
be deduced by comparing the low-energy behaviour of the $n$-particle
$S$-matrix elements to terms in the expansion in powers of $\Phi$ of
conventional principal chiral nonlinear sigma models.  Since $g$
has
an infinite expansion in powers of $\Phi$ it is complicated to carry
this out systematically beyond the lowest orders.  We will instead
consider  condition
for the vanishing of the one loop beta function that is the
lowest-order requirement of conformal invariance  will be
seen to be the familiar equations of motion for the principal
chiral models based on the group $G$.

\subsection{Low energy limit of S-matrix elements}
The low energy limit of the four point amplitude can  be obtained
from (\ref{finalfour}), (\ref{resfourone}) and (\ref{resfourtwo}) by
expanding the expressions in powers of $s$, $t$ and $u$.  This
corresponds to an expansion in powers of $\alpha'$.  The leading
terms come from (\ref{resfourone}), using $s+t+u=0$ and
\begin{equation}\label{expansion}
 \sin{\pi \alpha' s\over 2} + \sin{\pi \alpha' t\over 2} +\sin{\pi
\alpha' u \over 2} = -{\pi^3\over 16}\alpha^{\prime 3}stu +
o(\alpha^{\prime 5}),
\end{equation}
the contribution from (\ref{resfourone}) is,
\begin{equation}\label{a1lowenergy}
A^{(1)} =  - {\pi^2\over 2}  {\rm tr}(T^{a_1}T^{a_2}T^{a_3}T^{a_4}) \
u + {\rm perms}  +
o(\alpha^{\prime 2}),
\end{equation}
while the contribution from (\ref{resfourtwo}) is
\begin{eqnarray}\label{a2lowenergy}
A^{(2)} =  - {\pi^2\over 32}{\rm tr}(T^{a_1}T^{a_2}){\rm
tr}(T^{a_3}T^{a_4}) \alpha' tu + {\rm perms} +
o(\alpha^{\prime 3}).
\end{eqnarray}
Therefore, in the low-energy limit the dominant contribution to the
four-particle amplitude comes from $A^{(1)}$.   This is just the form
of
the amplitude expected for a nonlinear sigma model since it is
linear in the Mandelstam invariants which means that it comes from a
lagrangian with two space-time derivatives.  More precisely, the
amplitude is reproduced by an action  of the form,
\begin{equation}
\int d^Dx\left(\half{\rm tr}(\partial_\mu \Phi \partial^\mu \Phi) +
{\rm
tr} (\Phi
\partial_\mu \Phi \Phi \partial^\mu \Phi) - {\rm tr}(\Phi  \Phi
\partial_\mu\Phi \partial^\mu \Phi)\right),
\end{equation}
(where $\Phi \equiv \Phi^a T^a_{ij}$) which are the lowest-order
terms in the expansion of the action  (\ref{chiralact}) of the
principal
chiral model.  In order to verify that the leading $\alpha'$ terms in
the low-energy limit are identical to those of the principal chiral
model it is
necessary to compare amplitudes with  arbitrary numbers of external
Goldstone bosons with the amplitudes derived from the $\Phi$
expansion of the field theory lagrangian.  There are generally
intermediate massless poles in multi-particle amplitudes (a feature
that does not arise in the four-particle amplitude)
which must be subtracted explicitly before taking the low-energy
limit, which  complicates the procedure.

If the open-string photon is ignored the higher-order terms in the
$\alpha'$ expansion of the massless
scalar amplitudes give higher-derivative generalizations of the
nonlinear sigma model.  For example, the terms of order $\alpha'$
relative to the leading terms correspond to a model with four
derivatives.  However, the presence of the photon  affects the terms
beyond lowest-order
in $\alpha'$.

The observation that a single photon does not couple to any number of
massless Goldstone bosons is in accord with the fact that there is no
action linear in the abelian field strength $F_{[\mu\nu]}$ coupling
to a function of the group element, $g= e^{\Phi\cdot T}$.
The simplest candidate for a covariant coupling betwen photons and
the massless scalars is the linear combination of two terms,
\begin{equation}\label{photmod}
C\left(F_{[\mu\nu]} F^{[\nu \rho]}\ {\rm tr}
(g^{-1}\partial_\rho g\ g^{-1}\partial^\mu g) + \alpha F_{[\mu\nu]}
F^{[\nu \mu]}\ {\rm tr}
(g^{-1}\partial_\rho g\ g^{-1}\partial^\rho g)\right),
\end{equation}
which gives amplitudes with  two photons. The constant $\alpha$ is
undetermined in the low-energy field theory but may be determined by
comparison with the low-energy limit of the string theory amplitudes.
The low energy limit of the string amplitude is  obtained by
expanding
(\ref{photon3}) in the Mandelstam invariants,
\begin{equation}\label{photon4}
A_{\gamma\gamma} =\xi_1\cdot \xi_2  \Phi_3^a  \Phi_4^a
\alpha'\frac{\pi^2}{8}ut+\cdots+o({\alpha^\prime}^3).
\end{equation}
On the other hand the contribution of the two terms in
(\ref{photmod}) to the four-particle amplitude,
\begin{equation}\label{photfive}
A_{\gamma\gamma}  = \xi_1 \cdot \xi_2 \Phi_3^a  \Phi_4^a   C \left(
u^2+t^2 -2\alpha s^2\right) +\cdots.
\end{equation}
Comparison of this expression with (\ref{photon4}) uniquely
determines $\alpha =\half$ (and $C$ is an irrelevant overall
normalization).

\subsection{Equation of motion for principal chiral models}
Before discussing the condition for the vanishing of the one-loop
beta function  we shall review the form of the equations of motion of
the principal chiral models in terms of the field definitions that
arise in the boundary condensate.
 The variation of the lagrangian (\ref{chiralact}) under the
infinitesimal variation $\delta g=g\kappa $ is,
\begin{equation}\label{lert}
\delta L= {\rm tr}(\partial_\mu(g^{-1}\partial^\mu g)\kappa)
\end{equation}
so that the equation of motion has the form
\begin{equation}\label{motion}
\partial_\mu A^\mu = \partial_\mu(g^{-1}\partial^\mu g) = 0 ,
\end{equation}
where $A^\mu=g^{-1}\partial^\mu g$ is a flat non-abelian gauge
potential.   The equation of motion  (\ref{motion}) can be written as
an equation for $\Phi^a$ by making use of the identity,
\begin{equation}\label{soiu}
A^\mu=\sum_{n=0}^{\infty}{\frac{(-1)^n}{(n+1)!}
(Ad_{\Phi})^n\partial^\mu\Phi},
\end{equation}
where $Ad_CD=[C,D]$ (and $C,D$ are matrices) so that,
\begin{equation}\label{defad}
(Ad_C)^n D = [C, [C, [C, \cdots, [C,D]\cdots ] ] ]
\end{equation}
($\cdots$ indicates that there are $n$ $C$'s in the expression).
Applying another derivative gives the equation
of motion,
\begin{equation}\label{eqnmot}
\partial_\mu
A^\mu=\sum_{n=0}^{\infty}\frac{(-1)^n}{(n+1)!}\left\{(Ad_{\Phi})^n
\partial^2 \Phi+\sum_{k=0}^{n-1} (Ad_{\Phi})^k
(Ad_{\partial^\mu\Phi})(Ad_{\Phi})^{n-k-1}\partial^\mu\Phi\right\}
=0.
\end{equation}
These expressions are more familiar when written in terms of the
metric on the group manifold, $\gamma_{ab}(\Phi^c)$, with the action
density (\ref{chiralact}) given by $\half \gamma_{ab}\partial_\mu
\Phi^a \partial^\mu \Phi^b$, in which case the equation of motion
becomes,
\begin{equation}\label{groupeqn}
\gamma_{ab}\partial^2 \Phi^b + {\partial \gamma_{ab} \over \partial
\Phi_c}\partial_\mu \Phi^c \partial^\mu \Phi^b - \half {\partial
\gamma_{cb} \over \partial \Phi_a}  \partial_\mu \Phi^c \partial^\mu
\Phi^b = 0.
\end{equation}

\subsection{The one-loop beta function and the $\Phi$ equations of
motion}
The total action for a space-time dependent boundary condensate
is,
\begin{equation}\label{splus}
S=S_0+S_B,
\end{equation}
where $S_0$ is the bulk term and the boundary term is given  by,
\begin{equation}
S_B = \half  \int d\sigma \Phi^a(X(\sigma))  \left(J_L^a (\sigma) +
J_R^a (\sigma)\right),
\end{equation}
where $J_L^a(\sigma) = \psi_L^i(\sigma)T^a_{ij}\psi_L^j(\sigma)$ and
$J_R^a(\sigma) = \psi_R ^i
(\sigma)T^a_{ij}\psi_R ^j(\sigma)$.   We shall use a  background
field method in which the space-time coordinates $X^\mu$ are written
as the sum of a constant classical part ($\tilde X^\mu$) and a
quantum fluctuation ($\xi^\mu$),
\begin{equation}\label{fluct}
X(\sigma) =\tilde{X} + \xi (\sigma).
\end{equation}
The background-dependent coupling $\Phi(X(\sigma))$ is expanded in
powers of $\xi$,
\begin{equation}\label{expand}
\Phi^a(X)=\Phi^a(\tilde {X})+\partial_\mu\Phi^a
(\tilde{X})\xi^\mu+\frac{1}{2}\partial_\mu\partial_\nu
\Phi^a(\tilde{X})\xi^\mu\xi^\nu+o(\xi^3).
\end{equation}
Since each $\xi$
comes with a space-time derivative on the background field $\Phi$ and
we are interested in the low energy or  `small curvature' limit of
the
theory only a small number of derivatives need to be considered to
extract the leading $\alpha'$ dependence.
We shall keep only terms with up to two derivatives.   Using
the expansion (\ref{expand}) in the boundary action, $S_B$, gives
vertices with two fermion lines and arbitrary number of  $\xi$ lines.
 The  fermions and $\xi$ propagators evaluated between points on the
boundary are derived from the
bulk part of the action and satisfy the usual boundary conditions
(i.e., Neumann boundary conditions on $\xi$).

 We will calculate the one-loop counterterm  bilinear in  the fermion
fields that renormalizes the value of $\Phi$ in $S_B$. This has  the
form, $\int d\sigma \psi^i(\sigma) \Gamma_{ij} (\Phi)
\psi^j(\sigma)$.
If the quantum corrections preserve the property that the theory is
conformally invariant the overall coefficient of the logarithmically
divergent contributions   (which signal a breakdown of scale
invariance) must vanish which  means that
the renormalization group beta function of the scalar field must
vanish for the
background condensate to define  a   conformal quantum field theory.
This is equivalent to ensuring that the theory is BRST
invariant by imposing the condition,
\begin{equation}\label{brstinv}
Q_{BRST}e^{ \int d\sigma \Phi^a(X(\sigma)) J^a_L(\sigma)}|B\rangle
=0,
\end{equation}
where $Q_{BRST}$ is the bulk BRST charge.  In the Siegel
gauge this is equivalent to imposing the energy--momentum tensor
conditions, (\ref{viramodes}).  However, the fact that the boundary
state
satisfies this condition is not by itself sufficient to guarantee
BRST invariance due to possibility of an  anomaly that generates a
cosmological constant as argued by Fischler and Susskind
\cite{fisc1},\cite{pol2},\cite{callanx} (see section 5).

A priori,  infinitely many diagrams contribute to this divergence
(even to lowest non-trivial order in $\alpha'$, which involves a
contraction of two $\xi$'s)   because of the presence of the constant
term, $\Phi^a(\tilde {X})  \psi^i(\sigma)T^a_{ij}\psi^j(\sigma)$
which is   quadratic in the fermion fields.  It is sensible to
eliminate this
term in order to reduce the one-loop problem to the evaluation of a
single diagram.

It is convenient to use  the operator realization of the string path
integral  by quantizing  on equal $\tau$ slices  on the semi-infinite
cylinder  (where $\tau$ is the axis of the cylinder).   The currents
$K=(J_L - J_R)/2$ may be replaced by the left-movers $J_L$ using
the fact that the boundary state, $\mid B\rangle$, relates
right-movers to left-movers and the partition function is given by,
\begin{eqnarray}\label{exp1}
&&Z[\Phi] =  \langle P\mid \exp\int d\sigma  \Phi^a(X(\sigma))
K^a(\sigma)\mid B\rangle \nonumber\\
 &&= \langle P \mid \exp(\int
d\sigma\{\Phi^a(\tilde{X})+\partial_\mu\Phi^a
(\tilde{X})\xi^\mu+\half \partial_\mu\partial_\nu
\Phi^a(\tilde{X})\xi^\mu\xi^\nu+o(\xi^3)\} J^a_L(\sigma) )\mid
B\rangle,
\end{eqnarray}
where the presence of the state $\langle P|$ again allows for the
coupling of the
boundary to  an arbitrary closed-string state.

In order to evaluate  this expression in perturbation theory the
exponent is   written as a sum of two pieces, $C$ and $D$, where
\begin{equation}\label{defina}
 C =\int d\sigma\Phi^a(\tilde{X}) J_L^a (\sigma) =\Phi^a(\tilde
X)J_{L0}^a,
\end{equation}
is independent of $\xi$,
while the second involves all powers of $\xi$,
\begin{equation}\label{Dexponent}
D = \int d\sigma \left\{
\partial_\mu\Phi^a(\tilde{X})\xi^\mu+\frac{1}{2}
\partial_\mu\partial_\nu\Phi^a(\tilde{X})\xi^\mu\xi^\nu+o
(\xi^3)\right\}  J_L^a(\sigma).
\end{equation}
The term $C$  is the constant piece of  the exponent (\ref{exp1}) and
can be moved to the left  with the help of  Baker-Campbell-Hausdorf
formula,
\begin{equation}\label{BCH}
 e^{C+D}= e^C e^{\sum_{r=0,s=1}^\infty H_{r,s}(C,D)}.
\end{equation}
 Here the $H_{r,s}(C,D)$ are homogeneous polynomials of degree $r$ in
$C$ and $s$ in D which are  made out of multiple commutators of $C$
and $D$
The partition function may be written in terms of the $H_{n,m}$'s,
\begin{equation}\label{newz}
Z[\Phi] = \langle P \mid e^C  e^{\sum_{r=0,s=1}^\infty
H_{r,s}(C,D)}\mid
B\rangle.
\end{equation}
The factor of $e^C= e^{\Phi^a(\tilde X) J^a_{L0}}$ simply transforms
the  closed-string bra state and can be ignored.
 We will see that the one-loop logarithmic divergences come only from
terms in $H_{n,1}$ and $H_{n,2}$.  These terms can be obtained  by a
power series expansion of (\ref{BCH}) and are given (after some
algebra) by,
\begin{equation}\label{sum1}
H_{n,1}= \frac{(-1)^n}{(n+1)!}(Ad_C)^nD
\end{equation}
and
\begin{equation}\label{htwodef}
H_{n,2} =\frac{1}{2}\frac{(-1)^n}{(n+2)!}\sum_{k=0}^{n-1}
(Ad_C)^k Ad_D(Ad_C)^{n-k}D.
\end{equation}

As noted above we are interested in terms with up to two space-time
derivatives or equivalently two powers of the quantum fluctuations,
$\xi$. There are two contributions that we need to consider.
Firstly, $H_{n,1}$ with arbitrary $n$ contains terms of the form
$\partial_\mu\partial_\nu\Phi(\tilde{X})\xi^\mu\xi^\nu$ coming from
$D$.  The expression in (\ref{sum1}) can be simplified
by using the fact that $C$ is proportional to the zero mode of the
current so its commutation relations simply reduce to group theoretic
commutators, as can be seen from the general expression,
\begin{eqnarray}\label{adcd}
Ad_C \chi =\left[ \Phi^a(\hat{X})J^a_{L0},\int d\sigma
\chi^b(\sigma) J_L^b(\sigma)\right]\
&=& \int d\sigma   f^{abc} \Phi^a(\hat{X}) \chi^b(\sigma)J_L^c(\sigma
)\nonumber \\
&= & \int d\sigma   J^a_L (\sigma) Ad_\Phi \chi^a(\sigma)  ,
\end{eqnarray}
where $\chi^a(\sigma)$  is an arbitrary matrix-valued function and
the second line makes use of the definition of $Ad_C D^a$,
\begin{equation}\label{notation}
Ad_C D^a = (Ad_C D )_{ij} T^a_{ij}.
\end{equation}
 Denoting the
terms in the sum in (\ref{sum1}) (with C  defined by (\ref{defina})
and $D$ defined by (\ref{Dexponent})) that can lead to logarithmic
one-loop divergences by $H'_{n,1}$ we  find by iterating
(\ref{adcd}),
\begin{equation}\label{hone}
H'_{n,1}= \frac{(-1)^n}{(n+1)!}\int d\sigma  J_L^a(\sigma)
(Ad_{\Phi})^n\left\{\frac{1}{2}\partial_{\mu}\partial_{\nu}
\Phi^a(\tilde{X})\xi^\mu(\sigma)\xi^\nu(\sigma)\ + \partial_\mu
\Phi^a(\tilde X) \xi^\mu(\sigma) \right\}.
\end{equation}

The terms $H_{n,2}$ defined by (\ref{htwodef})  are quadratic in $D$
so they give rise to new terms that are quadratic in $\xi$ that arise
as bilinears in $D' =\int d \sigma \partial_\mu \Phi^a
J_L^a\xi^\mu$.
As before, terms of the form $(Ad_C)^k \chi$ in (\ref{htwodef}) are
given by the group theory that follows from (\ref{adcd}).  However,
the factors of the form $Ad_D(Ad_C)^{n-k}D$  involve a commutator of
two terms which are linear in both  $\xi$ and in the current
densities.   This commutator is given by,
\begin{eqnarray}\label{commone}
&&\left[\int d\sigma_1J_L^a(\sigma_1)\partial_\mu\Phi^a
\xi^\mu(\sigma_1),\int
d\sigma_2J_L^b(\sigma_2) (Ad_{\Phi})^{k_2}\partial_\nu\Phi^b
\xi^\nu(\sigma_2)\right]\nonumber\\
&&=\int d\sigma_1 \partial_\mu\Phi^a\xi^\mu(\sigma_1)\int
d\sigma_2(Ad_{\Phi})^{k_2}\partial_\nu\Phi^b\xi^\nu(\sigma_2)
\nonumber\\
&&\qquad\qquad \left\{f^{abc}  J_L^c
(\sigma_1)\delta(\sigma_1-\sigma_2) +\delta^{ab}\delta^\prime
(\sigma_1-\sigma_2)\right\}.
\end{eqnarray}
The terms on the right-hand side come from the equal time commutators
of the current densities, $J_L$.  The first term is,
\begin{equation}\label{termone}
\int d\sigma  J_L^a(\sigma) Ad_{\partial_\mu \Phi}(Ad_\Phi)^{k_2}
\partial_\nu\Phi^a \xi^\mu(\sigma)\xi^\nu(\sigma).
\end{equation}
while the $\delta'$ term  in (\ref{commone}) gives
$\int d\sigma \partial_\mu \Phi^a (Ad_\Phi)^{k_2}
\partial_\nu\Phi^b\delta^{ab}\xi^{[\mu}(\sigma) \partial_\sigma
\xi^{\nu]}(\sigma)$,
which will not lead to a logarithmic divergence.   As a result, the
terms in $H_{n,2}$ that can lead to one-loop divergences (denoted by
$H'_{n,2}$) are given by,
\begin{equation}\label{htwo}
H'_{n,2}=  \frac{(-1)^{n+1}}{2(n+2)!}\int
d\sigma J_L^a(\sigma)\sum_{k=0}^{n-1}(Ad_{\Phi})^k
Ad_{\partial_\mu\Phi}(Ad_{\Phi})^{n-k}\partial_\nu\Phi^a
\xi^\mu(\sigma) \xi^\nu(\sigma) .
\end{equation}

The modified action  given by $H_{n,1}+H_{n,2}$ may now be used to
calculate the logarithmic divergent graphs, given by the following
diagram,
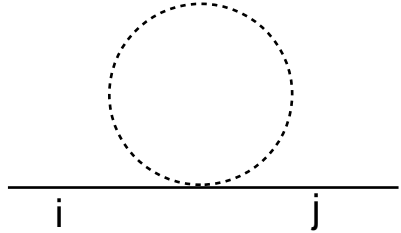
\null\vskip 0.2cm
\let\picnaturalsize=Y
\def\picsize{1.0in}
\def\picfilename{fpion1.eps}
\ifx\nopictures Y\else{\ifx\epsfloaded Y\else\input epsf \fi
\let\epsfloaded=Y
\centerline{\ifx\picnaturalsize N\epsfxsize \picsize\fi
\epsfbox{\picfilename}}}\fi

\noindent The vertex  is the $\xi\xi$ term in $H'_{n,1}$ or
$H'_{n,2}$ in (\ref{hone}) and (\ref{htwo}), respectively (the dashed
line indicates the $\xi$ propagator and the full lines $\psi $
propagators).  In evaluating the matrix element (\ref{exp1}) the
logarithmically divergent self-contraction  of
$\xi^\mu(\sigma)\xi^\nu(\sigma)$  at coincident points requires
regularization by a cut-off $\Lambda$ defined by
$\langle \xi^\mu(\sigma)\xi^\nu(\sigma) \rangle =
G^{\mu\nu}(\sigma,\sigma) =\delta^{\mu\nu} \ln \Lambda$. The
contributions from (\ref{hone}) and (\ref{htwo}) are,
\begin{equation}
a_1 = J^a_{L0}\sum_n \frac{(-1)^n}{2(n+1)!}  (Ad_{\Phi})^n
\partial_{\mu}\partial^{\mu}
\Phi^a \ln\Lambda.
\end{equation}
and
\begin{equation}\label{atwo}
a_2 = J^a_{L0}  \sum_n\frac{(-1)^{n+1}}{2(n+2)!}
\sum_{k=0}^{n-1}\int d\sigma (Ad_{\Phi})^k
Ad_{\partial_\mu\Phi}(Ad_{\Phi})^{n-k}\partial^\nu\Phi^a \ln
\Lambda.
\end{equation}

In this discussion the presence of the
term linear in $\xi$ in $H'_{n,1}$  (\ref{hone}) has so far been
ignored. This is a vertex with a
single $\xi$ that leads to the following one-loop diagram, which is
easily
seen to  vanish,

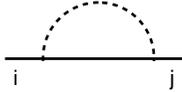
\let\picnaturalsize=Y
\def\picsize{1.0in}
\def\picfilename{fpion2.eps}
\ifx\nopictures Y\else{\ifx\epsfloaded Y\else\input epsf \fi
\let\epsfloaded=Y
\centerline{\ifx\picnaturalsize N\epsfxsize \picsize\fi
\epsfbox{\picfilename}}}\fi

The final result for the logarithmically divergent one-loop
counterterm, $\Gamma$,  is given by  $a_1 + a_2$. The  vanishing of
the beta function, $\beta^a=\Lambda^{-1}d\Gamma^a/d\Lambda=0$,  gives
the
equation of motion,
\begin{equation}\label{eqm}
\sum_n\left\{\frac{(-1)^n}{2(n+1)!}(Ad_{\Phi})^n\partial^2\Phi^a+
\frac{(-1)^{n+1}}{2(n+2)!}\sum_{k=0}^{n-1}(Ad_{\Phi})^k
Ad_{\partial_\mu\Phi}(Ad_{\Phi})^{n-k}\partial^\nu\Phi^a\right\}=0,
\end{equation} which is just the equation of motion of the principal
chiral model given in (\ref{eqnmot}), which is equivalent to
(\ref{groupeqn}).

As a  corollary we can now return to the discussion of
the nonlinear transformations described in the last section.
These nonlinear transformations can be deduced from the partition
function, (\ref{exp1}),  by considering the effect of shifting the
$\Phi$ field (now taken to satisfy the beta function equation)  in
the exponent of the partition function to
$\Phi'(X(\sigma)) =  \Phi^a(X(\sigma))  + \eta^a$ (with constant
$\eta$) and using (\ref{BCH}) with $C= \eta^a J_{Lo}^a$ and $D= \int
\Phi^a(X(\sigma)) J_L^a(\sigma)$.  For purely open-string processes,
in which $\langle P| = \langle 0|$ in (\ref{exp1}), the factor of
$e^C = e^{\eta^a
J_{L0}^a}$ is equivalent to unity on the external closed-string bra
state giving,
\begin{equation}\label{newzz}
Z[\Phi'] = \langle 0\mid \exp \int
d\sigma \sum_{r=0,s=1}^\infty H_{r,s}(C,D)\mid B\rangle
\end{equation}
(more generally,  the closed-string bra state transforms from
$\langle P|$ to $\langle P| e^{\eta^a J_{L0}^a}$).
We are interested in the terms up to linear order in $\eta$ in the
remaining exponent.
The term independent of $\eta$ is $H_{0,1} = D$ (all $H_{0,n}$ with
$n>1$ vanish) while $H_{1,n}$ are the terms linear in $\eta$
involving $n$ powers of $\Phi$. It is easy to verify that $H_{1,1} =
\half [C,D]$ and $H_{1,2} = -[D , [D, C]]/12$ so that the exponent of
(\ref{newzz}) can be written as $\int d\sigma \Phi^{\prime\prime a}
J^a_L(\sigma)$ where,
\begin{equation}\label{nonlin}
\Phi^{\prime\prime}(\sigma) =  {1\over 2}f^{abc} \eta^b
\Phi^c(X(\sigma))  +
{1\over 12}f^{abe} f^{ecd} \eta^c \Phi^b(X(\sigma)) \Phi^d(X(\sigma))
 +
o(\Phi^3).
\end{equation}
Thus, the theory is invariant under the nonlinear transformation
$\Phi \to \Phi' = \Phi + \eta -  \Phi^{\prime\prime}$.

\section{Mixing of open and closed strings}

The quantum loop
corrections to the tree-level theory generate the coupling between
open and closed strings.  It has long been appreciated  \cite{cs1}
that the mixing
of open and closed-string states in the Minkowski space theory gives
rise to a  mechanism whereby the closed-string antisymmetric tensor
potential ($B_{[\mu\nu]}$) gains a mass by
absorbing the massless neutral open-string vector state (with
potential $A_\mu$).  The effective action for the massive state can
be written
as
\begin{equation}\label{csdef}
\int d^Dx \left({1\over 6} dB^2 + {1\over 2} ({\rm g} B -
F)^2\right),
\end{equation}
where $F=dA$ and g is the open-string coupling constant.  The
modified gauge symmetry of this action is $\delta
B = d\xi$, $\delta A = d\Lambda + {\rm g} \xi$, where $\Lambda$ is a
scalar
and $\xi$ is a one-form.  This can be used to define a gauge in which
$A$ is eliminated, resulting in the lagrangian for a massive
antisymmetric potential.   In four dimensions the antisymmetric
tensor is equivalent to a scalar
state defined by $d\phi = *dB$ and (\ref{csdef}) displays  the usual
abelian Higgs mechanism (in the nonlinear limit in which the mass of
the Higgs particles are infinite).   The terms involving non-zero
powers of
g arise from a world-sheet with
the topology of a disk.   There is an interaction between the
closed-string state, $B$, on the interior of the disk and the
open-string state, $A$, on the boundary.  The $O({\rm g}^2)$ mass
term for $B$
comes from the coupling of two $B$ states on the interior of the
disk.   In addition, the dilaton  has a non-zero coupling to the
vacuum via its coupling to the boundary of the disk.  This generates
a non-zero cosmological constant by the Fischler--Susskind mechanism
\cite{fisc1}.  This phenomenon may be associated with  anomaly that
comes from contact terms that
arise in proving the BRST invariance of amplitudes
(\cite{pol2}\cite{callanx}). These contact terms can be ascribed to
the presence of a zero-momentum dilaton and trace of the graviton  in
the
closed-string channel that couples to the disk boundary.

Since the closed-string sector carrying
the gauge symmetry $G\times G$ at the enhanced symmetry points
couples linearly to the coset-space open-string states it is to be
expected  that (\ref{csdef}) should generalize to a non-abelian Higgs
mechanism.  The coset-space currents, $K^a$, correspond to a linear
combination of gauge bosons that should gain a mass by swallowing the
massless open-string scalar states.

The mixing of the coset-space pions with massless closed-string
gauge fields is seen from the non-vanishing of g$\langle {\rm
phys}|\oint d\sigma K^a(\sigma) e^{ik\cdot X(\sigma)} |B\rangle$
(where $\langle
{\rm phys}|$ is a physical closed-string gauge particle state and we
have explicitly included a factor of the open-string coupling to
illustrate that the pion fields are of order g).
Furthermore, the
fact that the theory contains the requisite gauge invariance can
be seen directly from the bulk world-sheet action, in which the
target-space gauge fields occur in the terms,
\begin{equation}\label{ferms}
\int d^2z \left(\psi_L(\partial + A_{L\mu}\cdot \partial X^\mu)\psi_L
+ \psi_R(\bar \partial + A_{R\mu}\cdot \bar \partial
X^\mu)\psi_R\right),
\end{equation}
where $A_{L\mu} = A_{L\mu}^a T^a_{ij}$ and $A_{R\mu} = A_{R\mu}^a
T^a_{ij}$ are the background gauge potentials.  In the absence of
boundaries conformal invariance requires these potentials to satisfy
the Yang--Mills field equations \cite{ft1},\cite{cfmp1},\cite{sen}.
The target-space theory is invariant under the gauge transformations,
\begin{equation}\label{gaugetrans}
 A_L \to A_L' = \rho_L^{-1} A_L \rho_L - \rho_L^{-1} d\rho_L, \qquad
A_R\to A_R' = \rho_R^{-1} A_R \rho_R - \rho_R^{-1} d\rho_R,
\end{equation}
where $\rho_L=\rho_L^aT^a_{ij}= e^{\Lambda_L} $  and
$\rho_R=\rho_R^aT^a_{ij}= e^{\Lambda_R}$ are the gauge parameters for
$G\times G$. This can be seen by substituting (\ref{gaugetrans}) in
(\ref{ferms}) and noting that the $\rho_L$ and $\rho_R$ dependence
can be eliminated by  redefining the world-sheet fermion fields by
\begin{equation}\label{gaugeym}
\psi_L\to \psi'_L = \rho_L \psi_L,\qquad  \psi_R\to \psi'_R =
\rho_R \psi_R.
\end{equation}
In the presence of boundaries this redefinition of the fermion fields
affects the boundary condensate because it changes the relation
between the left-moving and right-moving fermions  at the boundary.
The new boundary state must satisfy (dropping the primes),
\begin{equation}\label{newferm}
\left(\psi_R - i\rho_R\rho_L^{-1} \psi_L \right)|B\rangle_\rho = 0,
\end{equation}
where $\rho_L$ and $\rho_R$ are here functions of the boundary values
of $X(\sigma,\tau)$.  The boundary state satisfying these conditions
can be expressed in terms of the Lie-algebra valued  functions,
$\Lambda_L$ and $\Lambda_R$, by
\begin{equation}\label{modcon}
|B\rangle_\rho = \exp \int d\sigma\left( \psi_L^i(\sigma)
\Lambda_{Lij}(X(\sigma))  \psi_L^j(\sigma) - \psi_R^i (\sigma)
\Lambda_{Rij}(X(\sigma)) \psi_R^j(\sigma)
\right) |B\rangle.
\end{equation}
For infinitesimal $\Lambda = \Lambda_L -\Lambda_R$ it is clear that
the  dependence on the gauge parameters can be eliminated by shifting
the
scalar field in the boundary action,
\begin{equation}\label{scalshift}
\Phi\cdot T \to \Phi\cdot T + \Lambda_L - \Lambda_R ,
\end{equation}
which is a shift of the scalar fields in the coset $(G_L\times
G_R)/G_{diag}$.

This shift of the scalar fields is characteristic of the non-abelian
Higgs phenomenon.   The low-energy action for the massive gauge
fields embodying this gauge symmetry is,
\begin{equation}\label{higgsdef}
\int d^Dx\left({1\over 4} {\rm tr} F^2 + {{\rm g}^2\over 2} {\rm
tr}(A^K +{1\over {\rm g}^2}
g^{-1}dg)^2 \right),
\end{equation}
where $F$ is the non-abelian field strength for the
dim($G)+$dim($G$) gauge bosons in $G\times G$ and $A^K = (A_L-
A_R)/2$ are the components in the coset directions in the algebra.
The dependence on the open-string coupling constant is consistent
with the conventional normalization in which $\Phi$ (where
$g=e^\Phi$) is of order g. These scalar fields   can
be eliminated by a suitable gauge choice in the usual
manner.  The equations of motion that follow from this action should
be derived directly as the condition for conformal invariance of the
coupled bulk and boundary system.

The presence of the gauge field mass term in (\ref{higgsdef})
originates from the coupling of two massless vector states to the
disk, ${\rm g}^2\int dzdz'\langle V_\pm^{a \mu}(z) V_\pm^{b
\nu}(z')\rangle$
(where $V^{a\mu}_\pm =V^{a\mu} \pm \tilde V^{a\mu}$).
In the frame in which the disk is represented by a semi-infinite
cylinder this can be expressed as the matrix element,
\begin{equation}\label{vectwo}
A^{\mu\nu ab}_\pm={\rm g}^2\langle k\mid(\alpha_1^\mu J_{R1}^a\pm
\tilde{
\alpha}_1^\mu J_{L1}^a)\int_0^{2\pi} d\sigma \sum_{n,m}(\alpha_n^\nu
J^b_{Rm} \pm\tilde{ \alpha}^\nu_{-n}  J^a_{L-m})e^{-ik
X(\sigma)}e^{i(n-m)\sigma}\int  {dq \over q^{3-N}}\mid B \rangle,
\end{equation}
where the  length of the cylinder  (the radius of the disk) is the
one real modular parameter for this process.
This expression is divergent due to the tachyon in the cylinder
channel.  Regulating this in a gauge-invariant manner leads to mass
terms for the antisymmetric combinations of gauge fields (with vertex
operators
$V_-^{a\mu}$) while the gauge fields in the unbroken diagonal
group ($V_+^{a\mu}$) remain massless.

Thus, at the enhanced symmetry point half the gauge fields get
masses of order ${\rm g}^2$ by
absorbing the massless open-string scalars lying in the $G\times G/G$
coset space, leaving dim($G$) massless gauge fields in the diagonal
subgroup.

However, the complete analysis is complicated by the fact that the
massless closed-string scalars (the moduli) also couple to the disk
and may gain mass.
Their  masses are determined by the two-point functions,
\begin{equation}\label{twoscal}
\langle \Phi^{ab} \Phi^{cd}\rangle = \int \langle
J^a_L(z_1)J^b_R(\bar z_1) J^c_L(z_2) J^d_R(\bar z)\rangle_{disk},
\end{equation}
where $\Phi^{ab}$ denotes the closed-string scalars with vertex
operators $J^a_LJ^b_Re^{ikX}$.  Among several non-zero contributions
this expression possesses a divergence due to the coupling of a
massless state to the boundary of the disk which has the form,
\begin{equation}
\langle \Phi^{ab}\Phi^{cd}\Phi^{ef}\rangle_{sphere}\Delta\langle
\Phi^{ef}\rangle_{disk},
\end{equation}
where $\Delta \sim \int dq/q \sim \ln \Lambda$. The three-point
function on the sphere has the form $f^{ace}f^{bdf}$ and the
one-point function on the disk is proportional to $\delta^{ef}$ so
the divergence is due to the exchange of the trace $\Phi^{ee}$. This
is similar to the divergence associated with the massless dilaton
\cite{fisc1},\cite{pol2},\cite{callanx} and can be cancelled by
adding a cut-off dependent  condensate to the classical theory which
modifies the action by the addition of a term,
\begin{equation}\label{cutoff}
S=S_0+g\int d^2z J^e_L J^e_R \ln \Lambda,
\end{equation}
which corresponds to a non-conformal perturbation of the classical
theory.
There are additional mass terms in (\ref{twoscal})  which give finite
masses to $\Phi^{ee}$ as well as to the antisymmetric scalar fields,
$\Phi^{[ab]}$.

\section{Nonorientable world-sheets}

We now turn to a brief discussion of theories defined on
non-orientable world-sheets.  In such theories the closed-string
sector space of states is projected onto the subspace in which the
\lq twist' operator, $\Omega$ (that interchanges the left-moving and
right-moving spaces), has eigenvalue $+1$.  This means that
$(1-\Omega) |{\rm phys}\rangle =0$, where $|{\rm phys}\rangle$ is a
physical closed-string state.  This projection eliminates the
antisymmetric tensor field and half the $G\times G$ gauge fields,
leaving those in the diagonal $G$ subgroup.  Similarly dim($G$)
massless scalar moduli fields survive the projection.   The presence
of the operator
$(1+\Omega)$ in a closed-string one-loop amplitude generates the
Klein
bottle  in addition to the usual
toroidal world-sheet.  The open-string sector is restricted by the
requirement that $X(\sigma) = X(\pi - \sigma)$.  In the absence of
Chan--Paton factors this kills the states at every odd level, which
eliminates the massless photon state as well as the open-string
scalar states in the
Cartan subalgebra of the group $G$ for generic compactifications.
Loops of open string include a factor of $(1+ (-1)^N)$, which
generates non-orientable surfaces such as the M\"obius strip, in
addition to
the orientable ones.

At enhanced symmetry points there is an ambiguity in how the
open-string projection is defined.  If the generic projection is
applied at these special points there are dim$G - d$ massless
open-string scalars, parameterizing the space $G/U(1)^d$.  The
coupling of $n$ of these coset-space open-string scalar states to a
closed string is given by,
\begin{equation}\label{nonoriencoupling}
\langle {\rm phys}|K^{a_1}(\sigma_1)e^{ik_1\cdot X(\sigma_1)}\cdots
K^{a_n}(\sigma_n)e^{ik_n\cdot X(\sigma_n)} |B\rangle,
\end{equation}
where  both the bra and ket are
states on which $\Omega =1$.  Since $(1+\Omega)K^a(1+\Omega) = 0$
the coupling vanishes
unless $n$ is even so there is no coupling between a single
open-string scalar and a single closed string.   Therefore, pion
states do not mix with  closed-string states and the $G$
gauge bosons remain massless as do  the $G/U(1)^d$ open-string
scalars.

Alternatively, at the enhanced symmetry points the non-orientability
projection in the open-string sector is naturally generalized to $(1+
(-1)^{N+p^2/2})$, where $p^I$ lies in the (even) root lattice of $G$.
 In the
fermionic formulation this corresponds to treating all the fermions
equivalently (whereas the projection of the previous paragraph
distinguishes pairs of fermions that form the generators of the
Cartan subalgebra).  This projection is equivalent to allowing  the
center of mass, $x$, to wind around the around the compact
dimensions only an even number of times.  It eliminates all of the
massless open-string scalar
states.  In this case the only massless staes in the theory are in
the closed-string sector, namely, the graviton, dim($G$) massless
vectors, dim($G$) massless scalars and the dilaton.

The considerations of this paper can be generalized to include
non-trivial Chan--Paton symmetry factors.  Interesting work on the
compactification of anomaly-free open superstrings (\cite{sagnottia}
and
references cited therein) shows a rich breaking of the
Chan--Paton symmetry that is dependent on the background fields of
the
bulk theory.  These theories are naturally non-orientable.

\vskip 2.0cm

\noindent{\it Acknowledgments}\hfill\break
We thank the CERN TH Division for its hospitality and M.Gutperle
is grateful for financial support from the EPSRC and a Pannett
research studentship from Churchill College, Cambridge.

\end{document}